# ON A POSSIBLE INDICATOR OF HOMOGENEITY OF THE UNIVERSE

Nóra Fáy & B. Lukács

CRIP RMKI H-1525 Bp. 114. Pf. 49., Budapest, Hungary

lukacs@rmki.kfki.hu

**ABSTRACT**
A new method is described for recognising internal non-randomness in deep galaxy surveys.

## 1. INTRODUCTION

The present study is a report about the development of a *method* for elaborating deep *pencil-beam* (or, slightly more generally, narrow-cone) galaxy or quasar counts for *symmetry*. Therefore, after Chapter 2 no physical (cosmological) discussions will come, although the considerations behind will remain such. Our future purpose is to apply the method on various galaxy/quasar counts, and we cannot repeat every time the method described here. Also, in this way all steps are explicit and readers can criticise them properly.

The above-mentioned symmetry is always *Killing* symmetry (not a conformal one, see Chap. 2), and practically *homogeneity*, because pencil-beam counts do not give information about *isotropy*, and other Killing symmetries are seldom suggested in cosmology. Obviously one would expect (approximate) homogeneity in matter density in a homogeneous Universe, but in the last 20 years many observations yielded highly inhomogeneous galaxy/quasar distributions. This will be discussed in Chapter 2. Chapter 3 describes the method, whose final output is a single number called $R^2$. Low values would be hard to interpret at all, medium values indicate inhomogeneous matter distribution, so the lack of the symmetry, while high values show inhomogeneous short & middle scale matter distributions but on a Universe homogeneous on large scale. Then Chapters 4 & 5 determine which values of $R^2$ are low, middle and high. Chapter 6 is a brief outlook. 5 Appendices elaborate details which would disturb the readers in the main text.

## 2. THE PROBLEM

According to the "cosmologic principle", sometimes called the Copernican Principle, the geometry of the Universe should be such that no preferred point exist in it. This is clearly not an established fact but a *principle*; but it is seldom explicitly challenged. The argumentation for and against would be out of place in this study; we only want to emphasize that the question of large scale symmetries of the Universe is important. Now, because terrestrial observations show fair enough *isotropy*, and isotropy from *two* points would result in homogeneity + isotropy (Hawking & Ellis, 1973), *and* according to the Copernican Principle Earth cannot be a preferred point, geometries with full spatial symmetry (so SO(4), E(3) or SO(3,1)) are generally preferred when describing the Universe. However direct confirmation is rather difficult.

Of course, deep surveys for galaxies and/or quasars may help. Via the Einstein equation matter feels strongly the symmetries of geometry and vice versa. However again there are difficulties.

First, symmetries of matter do not necessarily coincide with symmetries of the geometry. We mention only single examples for both directions. *Vacuum* solutions (when the matter side of the Einstein equation is maximally symmetric, being 0 everywhere) demonstrate that the geometry can be *less* symmetric than matter. On the other hand, Lukács & Perjés (1977) showed that explicitly time-dependent Maxwell fields can coexist with time-independent geometry.

This difficulty is not serious for cosmological situations. Namely, the Universe well after radiation decoupling seems to be filled with dust:

$$R_{ik} - (1/2)g_{ik}R + \Lambda g_{ik} = -(8\pi G/c^4)T_{ik} \qquad (2.1)$$
$$T_{ik} = \rho u_i u_k \qquad (2.2)$$

where $R_{ik}$ is the *Ricci* tensor, formed from the *metric* tensor $g_{ik}$ (Eisenhardt, 1950).

Now Killing symmetry means the existence of vector field(s) $K^i$ along which translations do not alter the geometry, so $g_{ik}$. Therefore the respective Lie-derivative of $g_{ik}$ vanishes:

$$L_K g_{ik} = 0 \tag{2.3}$$

which, to make the equation explicit, reads as

$$K_{i;k} + K_{k;i} = 0 \tag{2.4}$$

the Killing equation. But if (2.3) holds, then by construction

$$L_K R_{ik} = 0 \tag{2.5}$$

too, so through (2.1) also

$$L_K T_{ik} = 0 \tag{2.6}$$

After some tedious but simple calculations (for some details see Lukács & Mészáros 1985) finally we get

$$L_K \rho = 0 \tag{2.7}$$
$$L_K u^i = 0 \tag{2.8}$$

For complete spatial symmetry this means in adapted coordinates

$$\rho = \rho(t) \tag{2.9}$$
$$u^i = \delta_0^i \tag{2.10}$$

(Robertson & Noonan, 1969). So density should be homogeneous, and the cosmological velocity field is complete rest.

The second difficulty is more serious. Galaxy counts are direct enough measurements, but they definitely show serious inhomogeneities. (Here we refer only to Broadhurst & al. (1990). A huge literature exists reporting inhomogeneities.) For first glance this would seem to indicate *no homogeneity*; but of course homogeneity can occur at scales higher than that of the observed inhomogeneities. Since this point does not really belong to statistics, which is the true topics of this paper, even the brief discussion of this point is relegated to Appendix A. Here we only refer 2 data for the scales.

Surely galaxy counts show large inhomogeneities up to cca. 150 Mpc. (In the seminal paper (Broadhurst & al. 1990) this was 128 Mpc.) And Mészáros (1997) observes transition from middle to large scale (cca. homogeneity) at 300 Mpc.

And then we are confronted with a string of practical difficulties beyond 300 Mpc ~ z=0.1. First, we are starting to lose the fainter galaxies in the counts, and, second, we cannot measure any *distance* on this scale, only redshifts. Even if we accept the cosmologic origin of redshifts (which we definitely do here), redshift z is *not* the comoving distance x. The latter, the primary quantity for Cosmology's viewpoint, can only be calculated from z, *in a model-dependent way*. Again, details are relegated to Appendix B, but obviously, a sample may seem fairly homogeneous on a large enough scale in one cosmologic model and fairly inhomogeneous in another, as demonstrated for example in (Horváth, Lukács & Paál 1992).

Summarizing: deep surveys are available. They are fairly exhaustive in the equatorial plane now up to z=0.2 (except directions where the galactic lens interferes) as the SLOAN database does (see Acknowledgment). Individual even deeper surveys are available in special directions. The surveys are either pencil-beam ones or can be made such, which is preferable in mathematical analyses to avoid the problems of recognition of involved structures. Then we get simply series

$$\{z_i\}$$

or

$$\{n_i; z_{i-1} < z < z_i\}$$

and the homogeneity of the underlying geometry of the Universe should be reflected by the homogeneity of the series. However:

1) All the series are inherently inhomogeneous up to $\Delta z \sim 0.05$ and may be inhomogeneous up to 0.1.

2) Longer observed series are not "stationary", i.e. there is no symmetry for shift, because i) pencil-beam surveys would give galaxy numbers $\sim r^2 \sim z^2$ for purely geometric reasons even for exactly homogeneous medium; and ii) with increasing z's the fainter galaxies are more and more lost. (In the SLOAN samples this seems to happen seriously already at z=0.15.)

3) The scales are not disjoint enough.

Therefore the recently available material cannot be used for the well-established homogeneity checks. However cosmology should not wait indefinitely for optimal samples. Indeed, non-biasing series $\{z_i\}$ beyond 600 Mpc, so in the scale of expected homogeneity, seem to belong to future generations of observational cosmology, see Appendix C. To give tentative answers to the questions of *large scale structure*, we have developed an unorthodox method of analysing series $\{z_i\}$ for underlying homogeneity/repetition, which we report in the remaining part of the present study. Please remember always the problems and difficulties listed in this Chapter 2 and Appendices A-C. Appendix D gives a solid state physics analogy to the problem. The method is handmade for definitely these difficulties and we do not advertise it for general use under different circumstances.

## 3. THE METHOD

We were looking for a single number characterising the homogeneity/repetition in a series $\{z_i\}$ (or $\{n_i; z_{i-1}<z<z_i\}$) *even if* the selection leading to the series was inhomogeneous. The method will be reported in separate steps; we give reasons for each step. While the reader may, of course, remain unconvinced about the optimal nature of any individual step, we think we can show that indeed the resulting quantity which we will call $R^2$ for reasons understood in due course does not mimic homogeneity/repetition if that is not present.

We present a method in such a way that readers of later papers, where $R^2$ values will be used for cosmologic arguments be able to repeat the calculations. Of course minor differences from rounding errors &c. may occur, but we will be definite and maybe clear.

**Step 0: The input series.** This work was not especially financially supported and neither of us is in observational astronomy. So all our inputs are taken from the literature or from other databases. Since no specific observational data will be reported in the present study, here we do not list the inputs. For any case, in some cases we could use individual redshifts of galaxies, so $\{z_i\}$, while in other cases first the z range had been divided into bins by the observers and then the numbers of galaxies in bins $\{n_i\}$ were reported. The first case does not need further preparation, but in the second case in later phases rearrangements into new bins are nontrivial. For this purpose we developed an approximate "stochastic" method. It will be described up to formulae in Appendix E; in the main text we only tell the idea behind. Namely, we have no right to assume homogeneity *within* bins, but using the neighbouring ones via interpolation we get information about the *trends* within a given bin. The method is fundamentally the same as used in numerical analysis for approximate numerical integrations.

We wanted to get narrow, more or less pencil-beam surveys, ideally meaning narrow *cones* or bi-cones ("North" and "South"). So the prepared input of the method is either a series of nonnegative integers $n_i$ for a finite amount of bins in redshift z, *or* a number of redshifts $z_i$ *in growing order*, at which galaxy was detected. In the second case possible Southern galaxies are clearly distinguished from possibly Northern ones even if the z values are near each other, because the mutual distances may be huge.

**Step 1: The spatial series to be analysed.** Now we prepare the series which will be used in the homogeneity analyses. For this the redshift values $z_i$ must be translated to distances. The definitions and formulae are given in App. B. However it is a matter for the main text that we do not know the *true* cosmologic model. Namely, even for maximally symmetric cosmologies filled purely with dust the past radii/scales of the Universe R(t), so the past distances $x_i$ and redshifts $z_i$ depend on the *cosmologic constant* $\Lambda$ which belongs to the Law of Gravity not to Matter, so it is not directly observable. In addition, the total mass density cannot be reliably observed because of the possibility of "dark matter": nonluminous, e.g. intergalactic, gases, exotic heavy particles, black holes, &c. So we remain with 2 unknown constants, which are generally taken as: $\Lambda$ normalised to a dimensionless number and $\Omega = (\rho/\rho_{crit})_{present}$, where $\rho_{crit}$ is the density for the k=0 Universe for $\Lambda$=0. (Here $\Lambda$ is a dimensionless constant, proportional to the constant $\Lambda$ in eq. (2.1). The value will not be used henceforth.)

However in these years it seems that

$$\Omega + \Lambda \approx 1 \qquad (3.1.1)$$

(again, proper references are beyond the scope of the present statistical study). So we remain on the line $\Omega + \Lambda \approx 1$. Since $\Omega>0$, and all observations give $\Omega$ well below 1, we take 11 $\Omega$ values

$$\{\Omega=0.1*n; \ 0\leq n\leq 1\} \qquad (3.1.2)$$

and

$$\Lambda = 1 - \Omega \qquad (3.1.3)$$

Thus for a specific $\Omega$ a distance $x_i$ will correspond to $z_i$:

$$z_i \rightarrow x_i(z_i;\Omega) \qquad (3.1.4)$$

The absolute scale in x is irrelevant here, but for definiteness' sake we accept here $H_o$ = 100 Mpcs/km.

We arrange the $\{x_i\}$ series (or rearrange the $\{n_i\}$ ones, see App. E) into bins, namely not less than 100 ones and not more than 300. For $\nu_{min}$ & $\nu_{max}$ the choices will be explained in due course; they seem useful. So we have 11*201 series:

$$\{n_i(\Omega); \ 0\leq\Omega\leq 1, \ 1\leq i\leq\nu, \ 100\leq\nu\leq 300\} \qquad (3.1.5)$$

**Step 2: Removing distance trends.** We will want to apply autocorrelation methods in due course. However the series $\{n_i\}$ contain strong trends for changing with distance which have nothing to do with the geometry of the Universe. Two such trends we have already mentioned, but surely there are others as well, of more or less unknown origin, as our experience (which will not be mentioned otherwise here) indicates. The 2 main reasons are as follows.

1) Pencil beam surveys happen in narrow cones. Then the respective surface of the survey at distance x is $\sim x^2$, so growing substantially.

2) With growing x, however, fainter galaxies are more and more lost.

This two reasons would give roughly

$$n_i \sim x^2 e^{-x/X} \qquad (3.2.1)$$

or something similar, where X might be taken empirically. While indeed such a correction helps, even (3.2.1) does not reflect the tendency in $\{n_i\}$. Therefore instead of theorising, we employ *moving averages*. Namely:

A) We *smoothen* the series in a range of 21 bins by forming the average

$$\tilde{n}_i = 21*n_i + \sum_{k=1}^{10} (21-2k)(n_{i-k}+n_{i+k})/141 \qquad (3.2.2)$$

Obviously we cannot use then 21 of the bins thereafter.

B) Then we form the quantity

$$y_i \approx n_i/\tilde{n}_i \qquad (3.2.3)$$

Then the series $\{y_i\}$ must be free of very long range tendencies (mainly consequences of observational selections) *in the amplitudes*, so standard techniques of autocorrelation can be applied. Of course, the "normalisation" equalises the amplitudes on ranges i>21 even if they were not equal, so we must not use arguments for homogeneity based on the long-range *amplitudes*.

C) The $\{y_i\}$ series is then can be symmetric for shifts; for middle range averages

$$<y> \approx 1 \qquad (3.2.4)$$

**Step 3: Autocorrelation.** *Structures* then can be seen from the autocorrelation of the series $\{yi\}$. Significant positive autocorrelation *with lag of k bins* indicates some repetition with scale length k or at least the presence of many objects with a characteristic length k. (The discussion of such repetitions belong to Cosmology Proper, so will not be discussed here, but think e.g. of "bubbles".) However slightly different quantities are called "autocorrelation" or "pair correlation". Here we use the following definition:

A) We have a series $y_i$, $1\leq i\leq n$. For the autocorrelation *with a lag of k bins* we formally form 2 series

$$y_{(1)i} = \{y_i; \ 1\leq i\leq n-k\} \qquad (3.3.1)$$
$$y_{(2)i} = \{y_i; \ k+1\leq i\leq n\} \qquad (3.3.2)$$

and then the respective *auto*correlation coefficient $r_k$ is the correlation coefficient of the above 2 series. So, and by definition:

$$r_k = <y_{(1)}*y_{(2)}> - <y_{(1)}><y_{(2)}>/\sigma_{(1)}*\sigma_{(2)} \qquad (3.3.3)$$

where

$$\sigma_{(A)} \approx (\langle y^2_{(A)} \rangle - \langle y_{(A)} \rangle^2)^{1/2} \qquad (3.3.4)$$

This $r_k$ has quite standard properties. It does not necessarily decrease with growing k, as the *pair correlation*, which is only the first term of the numerator, does. By construction it is always $-1 \leq r_k \leq +1$, so the meaning of *strong* correlation is clear.

For later steps we shall need the significance and mean deviation of $r_k$. For this let us change for *correlation* coefficients. For 2 different series again an r can be calculated, which is *exactly* 0 only if the two series are not only independent but also of infinitely long. For finite parts of lengths n of the *uncorrelated* series one gets

$$\langle r \rangle = 0 \qquad (3.3.5)$$
$$\langle r^2 \rangle - \langle r \rangle^2 = 1/(\nu-1) \qquad (3.3.6)$$

where for exact equality in (3.3.6) further assumptions are necessary, e.g. normal distributions within the series. However we may accept eq. (3.3.6) for first approximation Fisher & Yates 1953; Crow, Davis & Maxfield 1960). Also, with such assumptions, the distribution function of the variable r can be calculated which, however, strongly depends on the assumptions and will not be used in the followings. For any case one can expect that

$$r_k^2 > K^2/(n-k-1) \qquad (3.3.7)$$

with K = 2 (or 3) would mean that the hypothesis of uncorrelatedness is wrong; so such an autocorrelation really would indicate some *structure*.

Obviously if k is too large, the overlap area becomes too small, and statistical fluctuations seriously decrease the significance of the autocorrelation. Our experience shows that it is unreliable to go beyond

$$k_{max} = \nu(1-1/5) - 21 \qquad (3.3.8)$$

where the last term of course comes from the previous Subchapter.

**Step 4: Structure vs. noise.** We may assume short-range and longer-range autocorrelations. Then on larger scales the shortrange correlations appear as *noises*, even if this discrimination may not be fully objective. If the two ranges do not merge, then the autocorrelation shows the distinction. Namely, take a series $\{y_i\}$, long, stationary, &c., which is a sum of 2 terms, a structure-bearing $\{Y_i\}$ and a "noise" $\{\eta_i\}$. The respective autocorrelation coefficients are denoted r, R and ρ, and, "of course", we assume that $\{Y_i\}$ is uncorrelated with the noise $\{\eta_i\}$. So:

$$y_i = Y_i + \eta_i \qquad (3.4.1)$$

and then (3.3.3) leads to

$$r_k = (\sigma_Y^2 R_k + \sigma_\eta^2 \rho_k)/(\sigma_Y^2 + \sigma_\eta^2) \qquad (3.4.2)$$

where $\sigma_Y$ and $\sigma_\eta$ stand for the mean deviations *within* the subseries. So with growing k the direct influence of the noise goes away. However still the value of $r_k$ can diminish seriously if the amplitudes of "signal" and "noise" are comparable. To make this more explicit, let us introduce the "quality factor" Q

$$Q = \sigma_Y/\sigma_\eta \qquad (3.4.3)$$

Then

$$r_k = (R_k + \rho_k/Q^2)/(1+1/Q^2) \qquad (3.4.4)$$

So indeed the noise does not disturb us too much if, say, Q>2.

Still, the signal will be washed out if Q<1. However if there is repetition of the "signal" then it occurs a few times in the series $\{y_i\}$. If there is no repetition, clearly the result is negative anyway, so we may follow with expecting repetitions.

**Step 5: "Folding" of the autocorrelation.** Take the autocorrelation coefficient as a new series, of length (3.3.8), then take a folding length $\Delta < (\nu-21)/10$ (the divisor comes purely from experience), so divide it into N "leaves" of lengths $\Delta$, ignore for a while the first leave and the remainder at the end, and average as

$$\underline{r}_k(\Delta) = \Sigma_{L=2}^{N} r_{k+L\Delta} \qquad (3.5.1)$$

Then the repetition length is either near to the folding length $\Delta$, or not. If not, the averaging washes away the signal. However if the two lengths are close, the quasiperiodic peaks of the autocorrelation $r_k$ get (almost)

"above each other" after folding and do *not* average away. On the other hand noises do. So *in optimal case* the signal/noise ratio improves with $N^{1/2}$.

We try with each possible folding lengths, and tell that the "best folding" is when $\max(r_k)$ for k, which, of course, does depend on the folding length Δ becomes maximal over all Δ's. This quantity is called R(ν):

$$R(\nu) = \max_\Delta(\max_k(r_k(\nu))) \qquad (3.5.2)$$

**Step 6: Averaging on the manifold of bins.** Obviously the proper positioning of signals after folding is sensitive on bin number. Even if there is a (quasi)periodic repetition, generally the repetition length will not contain integer number of bins. So for some bin sizes $r_k$ and so $r_k$ may be substantial, while with other bin sizes it will not significantly differ from 0. While this is obviously true and can be demonstrated, it would be dangerously subjective to look for "good" bin sizes. Our criterion, instead, will be objective even if in some cases the signal might be washed away; when something remains, we can be sure that the signal is not our artifact.

As we mentioned at (3.1.5), the method is parallelly applied on 201 series, according to the 201 rearrangements of bins. Now we *average* the *squares* of the maxima:

$$R^2 = (1/201)\sum_{\nu=100}^{300} R^2(\nu)/\{1/\nu\} \qquad (3.6.1)$$

and we are ready. This averaging has the following reason:

R(ν) is an autocorrelation-like quantity in a calculation when the original autocorrelation was calculated from cca. ν points. (It was rather ν - k, but the k index is already dumb, and it is generally «ν.) Thus according to (3.3.6) one could expect an accidental value of R(ν) in the order of $(\nu)^{-1/2}$. So $R^2$ of (3.6.1) is a $\chi^2$-type quantity, with all terms behind the summation expected in the order of 1 and positive.»

Therefore $R^2$ cannot be *much below* <1. The meaning of "much below" will be attacked in the next two Chapters. If $R^2$ is "in the neighbourhood of 1", we do not see the signal; maybe the series were uncorrelated, maybe we washed away the signal. However if $R^2$ is "significantly higher", then some autocorrelations were present, the original series contained some repetition-like structures.

**Step 7: The "true" cosmology.** Still we have 11 values for $R^2$ according to the 11 Ω values considered. We do not go into details here because the possible consequences are purely cosmological. We only note that if the $R^2(\Omega)$ values do not differ significantly, then no consequence is possible. If, however, for some Ω range $R^2$ is significantly higher than at others, then one may argue for the respective Ω's. Namely, in a symmetric Universe, as told earlier, one does not expect "distorted repetitions", so e.g. bigger and bigger structures at greater and greater distances. But an incorrect Ω value just does mimic this. So if "regularity" is significantly better for some Ω, that is an argument that that particular Ω is the true one. Of course it is another question, *how strong* is the argument, and this point is outside of the scope of the present study.

Obviously if we know the significance levels belonging to $R^2$, we can decide if $R^2$'s for different Ω's do significantly differ or not.

The brief next Chapter will give some theoretical considerations for $R^2$. While they are based on statistics, $R^2$ has been calculated in many steps, so it would rather be hopeless to *calculate* the expectation value, mean deviation and significance levels of $R^2$. That will be made by "Monte-Carlo" methods later. However then it will turn out that even the simplified Chapter 4 is almost correct; and the theoretical approach, on the other side, will corroborate the numerical simulations.

## 4. THEORETICAL CONSIDERATIONS FOR THE DISTRIBUTION OF $R^2$

$R^2$ of eq. (3.6.1) is a $\chi^2$-type quantity by construction. Also, it is such a quantity in physical sense because it is the average of the squared deviations of the measured and theoretical values, divided by the squared mean deviations.

Namely, consider the *null hypothesis* that the series $\{y_i\}$ did *not* contain any autocorrelation except short-range ones ("noises"), which automatically averaged away in the process leading to $R^2$, if not earlier, then at Subchapter 3.5. Then the remaining autocorrelations would be 0, but because of the finite lengths of the series, as fluctuations $R^2(\nu) \approx (\nu)^{-1}$ results would occur. So the prediction of the null hypothesis is $R^2(\nu)=0$ but with $(\nu)^{-1}$ squared mean deviation, and indeed all members of the summation have expectation values $\approx 1$.

However observe that not all members of the summation are independent. It is clear that if you have a bin arrangement of say 317 bins, and then you rearrange it into 318, the two results (of any type) are highly correlated. Our supervision of the process suggests that only 1 rearrangement amongst 5 or 8 can be considered "essentially independent".

For any case, the theoretical expectation (Jánossy 1965) is that

$$\langle R^2 \rangle \approx 1 \qquad (4.1)$$
$$\langle (R^2)^2 \rangle - \langle R^2 \rangle^2 \approx 2/(m-1) \qquad (4.2)$$

where m is "the number of effectively independent points" and the distribution of $R^2$ is approximately a $\chi^2$-distribution with a degree of freedom m, where

$$m \sim 201/(5 \text{ or } 8) \qquad (4.3)$$

So $25 < m < 40$ and then $0.28 > \sigma_{R^2} > 0.22$

At m=30 therefore, the probability that $R^2$ from one sample is above cca. 1.6 *without really any not short-range structure* would be below 2.5% We stop here, because the next Chapter will show more detailed numerical simulations.

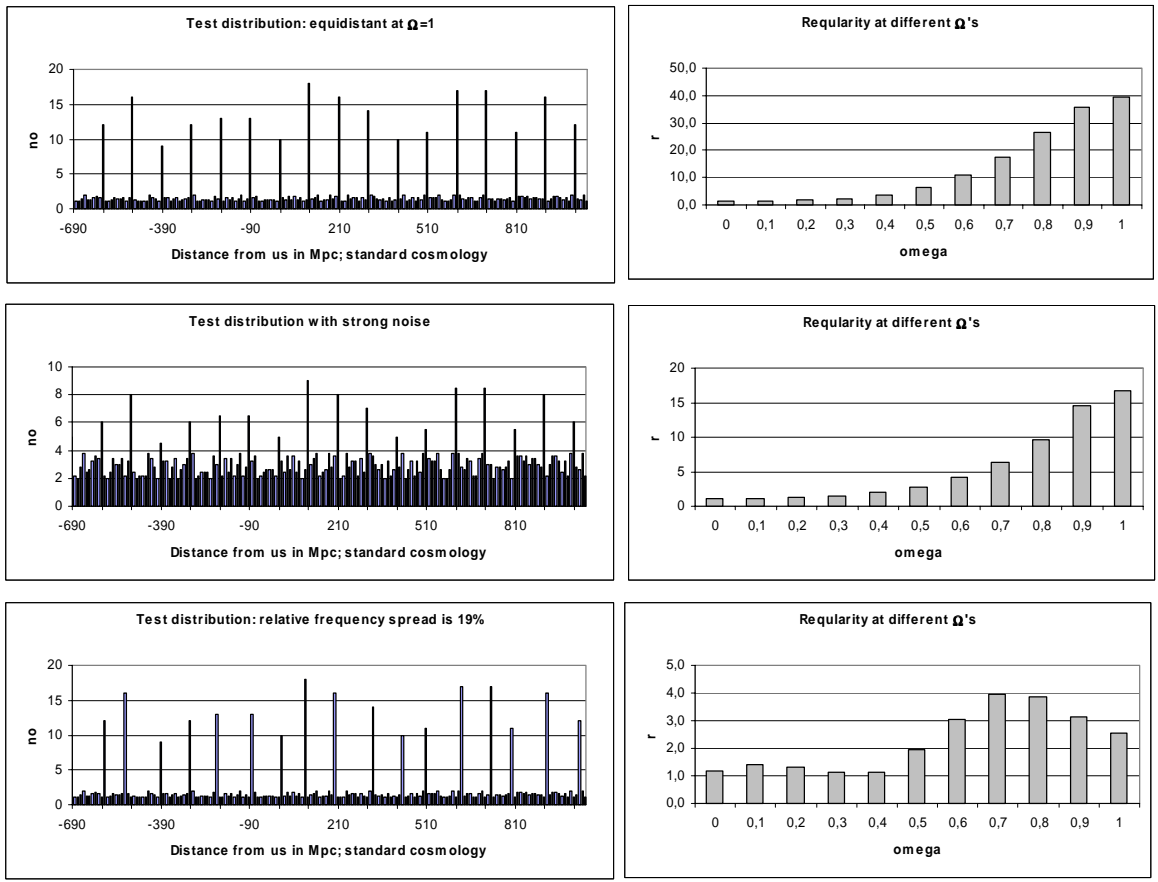

Fig. 1: Semi-random sequences for simulation. Above: equidistant (for $\Omega=1$), with amplitude modulation and small noise. The resulting $R^2$ goes up as high as 40, at $\Omega=1$. Middle: as above, but with a much stronger noise. $R^2$ is still 16 at $\Omega=1$, which is still maximum. Below: Frequency is also showing a spread, as high as 19 % relative. $R^2$ is still up to 4, but now the maximum is at $\Omega=0.7$. See the text below.

Observe that this is the reason for $\nu_{max} - \nu_{min} \sim 200$. A $\chi^2$–distribution with m far below 25 might have unfortunate statistical properties.

## 5. SIGNIFICANCE LIMITS FROM NUMERICAL SIMULATIONS

We repeated the procedure described in Chapter 3 for artificial series, generated by the combinations of deterministic and random processes. First let us mention briefly some results for artificial series with built-in repetitions. The original series was prepared always at $\Omega=1$, so it had the maximal structure (if any) there.

As we told in Chapter 2, there are the geometric factors (increasing surfaces in the pencil beam with increasing distances), and the increasing loss of the fainter galaxies with increasing distance. The result of these two phenomena is a peak in $\{ñ_i\}$ somewhere at $z=0.15$. We simulated this phenomenon with a parabola 0 at $z=0$ and at the maximum value; and according to our main purposes, we added up two such parabolas, on the two sides "of our telescope" (so $z=0$). (See an example at Fig. 3 later.) This completely deterministic factor is then multiplied with a semi-random component.

First we manufactured peaks at equidistant points with equal heights, plus a "noise" at 1/15 average level. $R^2$ went up so high as 167 for $\Omega=1$, and dropped rather fast with decreasing $\Omega$. As we expected, the method is strong in recognising equidistant cosmologic structures, but such a regular pattern, of course, cannot be expected.

Then we kept equidistancy, however applied a randomised height on the peaks, and a true random noise with average 1/13. (Fig. 1.) Still $R^2$ went up to 40 at $\Omega=1$, again decaying rapidly with decreasing $\Omega$. Enhancing the noise to ½ of the signal still $R^2$ is 16 at $\Omega=1$

The third level of randomness was to permit moderate random changes in the peak distances, in addition to the previous two variations. Now, since this *frequency modulation* is random, one may characterise it simply by a dimensionless constant which is the relative mean deviation of the "wavelength", so $\sigma_\lambda/\lambda$. It was practically impossible to go above cca. $\sigma_\lambda/\lambda=0.5$, because then peaks may "collide"; however Fig. 2 shows 288 points for various $\sigma_\lambda/\lambda$'s; now for the highest values on the $\Omega$ manifold (which is, however, generally $\Omega \approx 1$ if $\sigma_\lambda/\lambda$ is not too high). One can see that the extrapolated curve indeed is in the vicinity of $R^2=1$ without regularities in peak positions, but this is not the best way to calculate this limit.

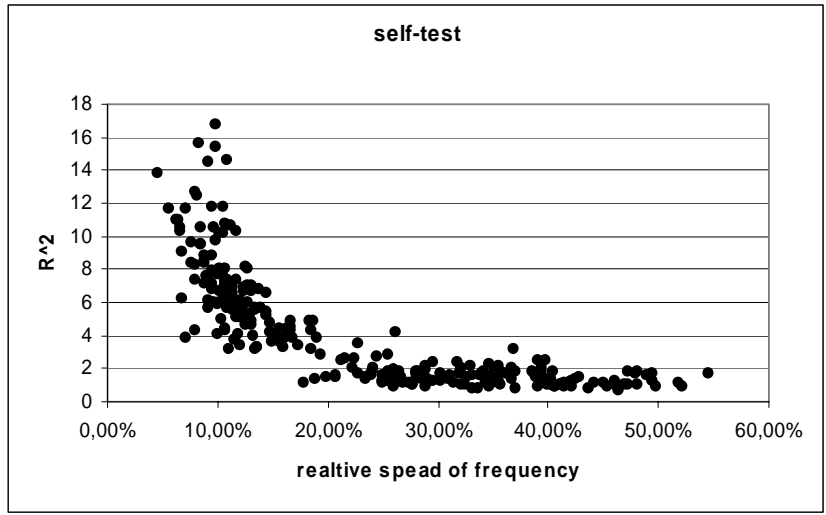

Fig. 2: $R^2$ vs. relative frequency spread for 288 simulated sequences. By construction it would be very difficult to exceed 0.5 in the relative frequency spread, but extrapolations suggest $R^2=1$ or somewhat slightly lower at infinite spread.

Now we turn to the numerical simulation of the null hypothesis of: galaxy samples from pencil-beam deep surveys without medium or long-range repetitions. This involves two factors.

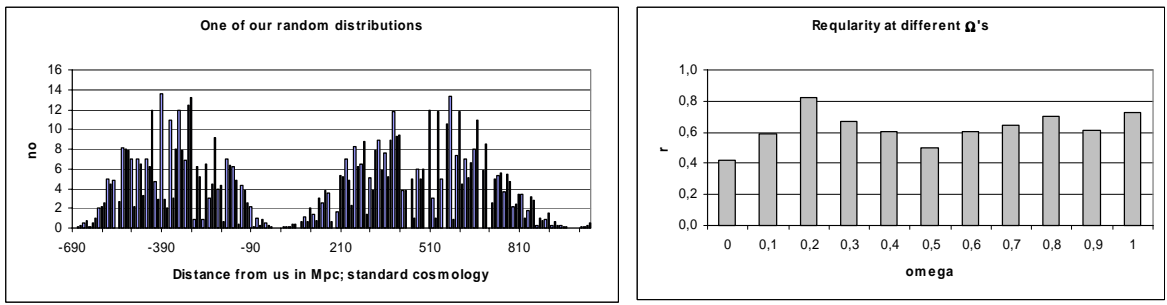

Fig. 3: A "random" series: the deterministic function simulating geometry + apparatus sensitivity times a series of random numbers between 0 and 10 in 176 sites.

This completely deterministic factor was then multiplied with a completely random series, thus resulting in the "deep survey samples without internal structures". But we must not take all 11 values for different $\Omega$'s, since these values are seriously correlated. Instead, the next Figure gives the results for $\Omega = 0.5$; the other 10 distributions are not much different.

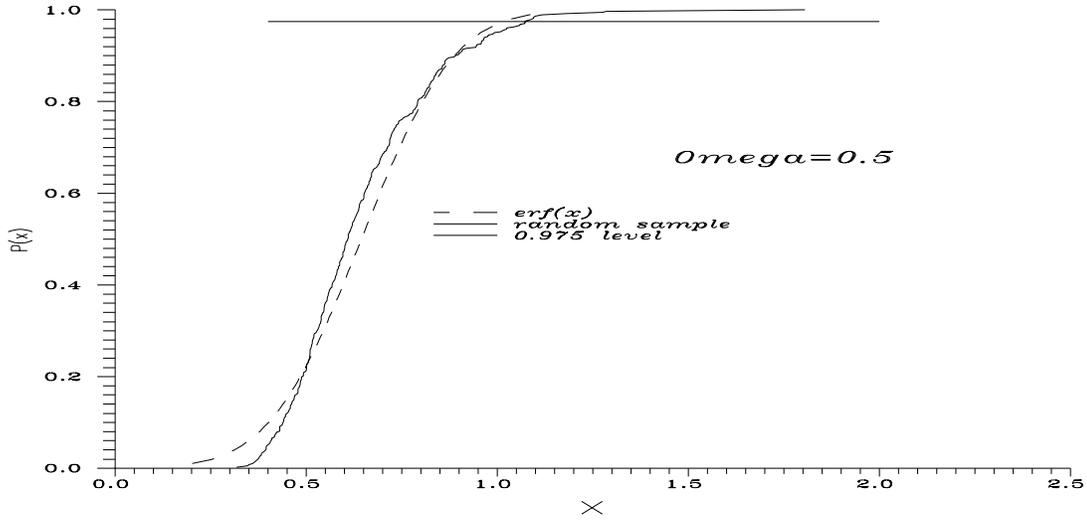

Fig. 4: The cumulative distribution of R$^2$ for "random" sequences for Ω=0.5, from 382 runs, compared with the Gauss distribution with the same first and second momenta. There are 2.5 % of the distribution above the horizontal line.

Fig. 4 shows the result of 382 simulations for cumulative probabilities, compared to the erf function which would be that for a Gaussian distribution, with the same first and central second momenta. As it can be seen, the difference is not great, but definitely the R$^2$ distribution differs in a weaker left tail and a stronger right one as indeed expected for $\chi^2$ distributions. For the momenta and errors we can proceed as follows. First momentum is unbiased estimate for the expectation value, and second central momentum, the square of standard deviation, can be obtained from the averages of the empirical squared deviation, but this is unbiased only if the divisor is n-1, not n. Then for distributions not far from Gauss the stochastic error of the expectation value is

$$\delta\langle x\rangle \approx \sigma/\sqrt{n} \tag{5.1}$$

For the stochastic error of $\sigma$ we cannot calculate an exact result, being the degree of freedom of the $\chi^2$ distribution unknown. However the distribution is close to Gaussian, and we can calculate exactly for a Gaussian in maximum likelihood (Jánossy 1965), obtaining

$$\delta\sigma \approx \sigma/\sqrt{2n} \tag{5.2}$$

where, by the construction of the maximum likelihood method n under the square root may well be n-1 or even n-3, but that would make small enough difference.

| Λ | Ω | $\langle R^2 \rangle$ | error δ | $\sigma_{R2}$ | error δ |
|---|---|---|---|---|---|
| 1 | 0 | 0.6614 | 0.0094 | 0.1816 | 0.0062 |
| 0.9 | 0.1 | 0.6625 | 0.0089 | 0.1718 | 0.0058 |
| 0.8 | 0.2 | 0.6556 | 0.0092 | 0.1766 | 0.0060 |
| 0.7 | 0.3 | 0.6497 | 0.0090 | 0.1729 | 0.0059 |
| 0.6 | 0.4 | 0.6532 | 0.0103 | 0.1986 | 0.0068 |
| 0.5 | 0.5 | 0.6431 | 0.0099 | 0.1909 | 0.0065 |
| 0.4 | 0.6 | 0.6325 | 0.0098 | 0.1885 | 0.0064 |
| 0.3 | 0.7 | 0.6219 | 0.0091 | 0.1764 | 0.0060 |
| 0.2 | 0.8 | 0.6105 | 0.0094 | 0.1816 | 0.0062 |
| 0.1 | 0.9 | 0.6102 | 0.0101 | 0.1939 | 0.0066 |
| 0 | 1 | 0.6024 | 0.0102 | 0.1974 | 0.0067 |

Table 1: Expectation values, mean deviations and their standard errors of $R^2$ as functions of Ω.

It is best to use two-sided significance boundaries. We remain at 95% level, so, from the random series, determine for every Ω a lower boundary below which there is 2.5% of the $R^2$ values, and an upper one above which is 2.5%. We of course will rather concentrate on the upper one. The results can be seen in Table 2.

The second and third columns give the significance boundaries for the $R^2$ of an individual sample. These boundaries are interpolated from the 382 random $R^2$ values, so "9.55" values are below and above. Column 4 is a number characterising the "right tail" of the distributions. For Gaussian ones the upper boundary would be at $\langle R^2 \rangle + 2\sigma_{R2}$. However $\chi^2$ distributions are asymmetric, their right tails are stronger, so now the upper boundary will be at $\langle R^2 \rangle + c_{cor}\sigma_{R2}$. Indeed, the factor is always >2, but only moderately. For a $\chi^2$ distribution with a degree of freedom 25 we would expect cca. $c_{cor}$ =2.4, and the result is around that value, indeed.

Columns 5 & 6 of Table 2 must not be confused with Columns 2 & 3. They are calculated from the errors of $R^2$. If we have a manifold of samples with N>>1, then the averaged $R^2$ value must be between the two significance boundaries if the samples do not show the regularity for which the method is sensitive. If not, the (by construction) random test distribution is distinguishable from the actual ones, so the latter ones are not random. Of course a single sample can as much vary as one member of the random sample, and this is the reason that the strip 2&3 is much wider; the factor is ≈√382.

| Ω | Two-sided significance | boundaries for $R^2$, 95% | Factor $c_{cor}$ | $R^2 - 2\delta R^2$ | $R^2 + 2\delta R^2$ |
|---|---|---|---|---|---|
| 0 | 0.407 | 1.07 | 2.249893 | 0.640183 | 0.682556 |
| 0.1 | 0.4 | 1.12 | 2.66341 | 0.638771 | 0.686213 |
| 0.2 | 0.393 | 1.07 | 2.346107 | 0.634167 | 0.677133 |
| 0.3 | 0.381 | 1.073 | 2.448706 | 0.627771 | 0.671663 |
| 0.4 | 0.365 | 1.111 | 2.305347 | 0.629443 | 0.676917 |
| 0.5 | 0.378 | 1.074 | 2.257147 | 0.620723 | 0.665409 |
| 0.6 | 0.368 | 1.123 | 2.603054 | 0.607017 | 0.657884 |
| 0.7 | 0.355 | 1.072 | 2.551069 | 0.59859 | 0.645261 |
| 0.8 | 0.348 | 1.033 | 2.326363 | 0.588646 | 0.632452 |
| 0.9 | 0.339 | 1.047 | 2.2532 | 0.587516 | 0.632814 |
| 1 | 0.343 | 1.003 | 2.029086 | 0.581595 | 0.623139 |

Table 2: Some significance boundaries for $R^2$.

So according a real galaxy survey giving $R^2$>1.123 the probability for accidental "regularity" is not more than 0.025 for *any* Ω, the usual 2σ significance level of Gaussian distributions.

The method seems applicable for *quasar* samples as well. Of course, for quasars the "double peak" (the interplay of geometry and loss of fainter sources) is at higher z's, so it is best to repeat the simulations leading to Tables 1 & 2.

And this point is proper to try to find the "best $\nu_{max}$". At the end of Chap. 4 we already formulated arguments for $\nu_{max} - \nu_{min} \sim 200$, and then we must not go with $\nu_{max}$ downward below 300. However we can go upward. We calculated the averages, mean deviations and errors repeating the calculations between 200 and 400 bins. The result is interesting: the higher calculations give "nicer" results, but there the method is not so sensitive. It is easy to summarize the results.

Between $\nu=200$ and 400 we get averages, mean deviations and errors independent of $\Omega$ in the sense that there is a common $<R^2>$ and $\delta_{<R^2>}$ so that for any particular $<R^2>(\Omega)$ $<R^2>-2\delta_{<R^2>}$ $< <R^2>(\Omega) < <R^2>+2\delta_{<R^2>}$, the averages are statistically indistinguishable. These common values are

$<R^2> = 0.9195 \pm 0.0080$ (5.3)

$\sigma_{R^2} = 0.1574 \pm 0.0057$ (5.4)

Compared with the approximate but theoretically based results of the previous Chapter ($R^2 \approx 1$, $\sigma_{R^2} \approx (0.22-0.28)$) the agreement is surprisingly good. However the averages of the random samples are higher, so indeed between 200 and 400 bins the method could separate less the random and non-random samples than between 100 and 300.

If this is the situation, the difference should be directly seen by recalculating $R^2$'s for the "half-random" series with checked frequency spread, so Fig. 2 as Fig. 2':

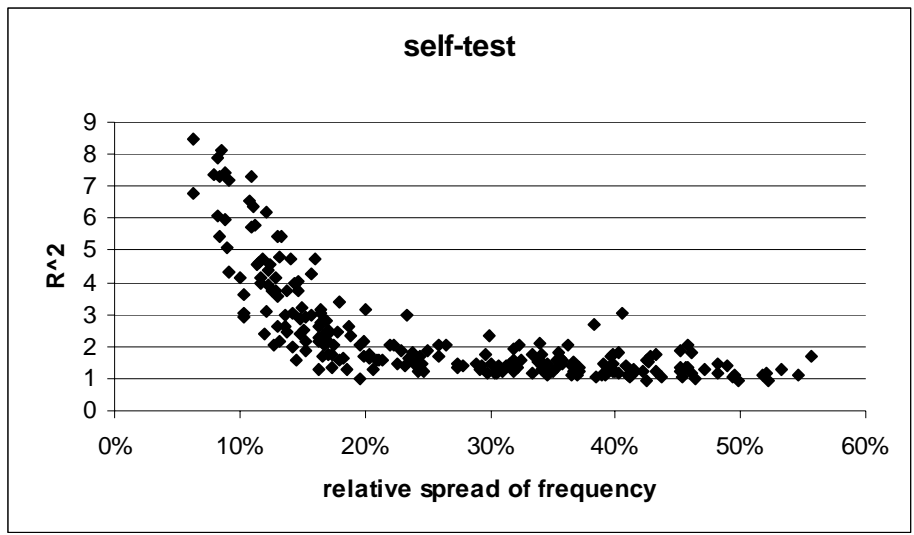

Fig. 2': As Fig. 2, but between 200 and 400 bins; 223 simulated sequences. The points lie higher for large spread (almost random pattern) and lower for small spread (very regular pattern). $R^2$ is less dependent on the frequency spread, so the method is less sensitive there.

We have some guesses, feelings and numerical experience to explain why the method is more sensitive between 100 and 300 bins; but in first approach it is best to regard this as a numerical result.

## 6. OUTLOOK

Our goal to publish the method separately from evaluating true samples was to guarantee reproducibility. Each step is given in details, so we think that everybody can now repeat the calculations, and then, up to statistical errors which were discussed, the results must be the same. Everybody checking is welcome.

As we told in Chap. 5, if one gets $R^2$ *significantly greater* than the expectation value in Table 1, then the sample is not random; what is more, it must contain *some* repetition. Generally the test does not show, *what kind* of repetition; but Step 5 in Chap. 3 is constructed in such a way that monotonously changing "repetition" lengths are not detected. We do not know how to interpret a situation when is *significantly lower* than the expectation value; so far we have not met such a case, and you can find a (not too exhaustive) discussion of the problem in Jánossy (1965). When the value does not differ *significantly* from the expectation value, the test is indecisive.

Now one must be careful with the meaning of *significantly*. First, there is the question of *significance level*. In Table 2 we gave the results formally for 95%; using Table 1 and some general knowledge about the behaviour of $\chi^2$ tails with degree of freedom $\approx 25$ one can calculate for other levels as well. However observe that this is a two-sided significance level. So if $R^2$ is above the upper boundary, then that is already 97.5%. In a more definite formulation, the chance to get such a high value by fluctuation is only 2.5%.

Still there remains *which* of the two intervals in Table 2 is to be used. The interval of Columns 2 & 3 is calculated via the mean deviation. Being the null hypothesis *no regularity*, such a sample must resemble the random sequences, so the significance interval indeed must be calculated via the mean deviation. However when averages of many $R^2$ 's of samples are discussed (for any purpose), then the measure of difference is the *error* of the expectation value, and then Columns 5&6 are relevant.

We are just applying this method on some SLOAN galaxy samples and also on some deep individual samples such as the Galactic Pole (Koo & al. 1993), SA68-anti-SA68 (Koo & al. 1993), Hercules (Munn & al. 1997), Lynx (Munn & al. 1997), SA57 (Munn & al. 1997) and Sculptor (Bellanger & de Lapparent 1995) ones. The results will be published in due course.

**ACKNOWLEDGEMENT**

We acknowledge some use of the SSDS (SLOAN) galaxy survey data, in which case we have to include the following information.

Funding for the Sloan Digital Sky Survey (SSDS) has been provided by the Alfred P. Sloan Foundation, the Participating Institutions, the National Aeronautics and Space Administration, the National Science Foundation, the U.S. Department of Energy, the Japanese Monbukagakusho, and the Max Planck Society. The SDSS Web site is http://www.sdss.org.

The SDSS is managed by the Astrophysical Research Consortium (ARC) for the Participating Institutions. The Participating Institutions are The University of Chicago, Fermilab, the Institute for Advanced Study, the Japan Participating Group, The Johns Hopkins University, Los Alamos National Laboratory, the Max-Planck-Institute for Astronomy (MPIA), the Max-Planck-Institute for Astrophysics (MPA), New Mexico State University, University of Pittsburgh, Princeton University, The United States Naval Observatory, and the University of Washington.**APPENDIX A: ON MIDDLE SCALE INHOMOGENEITIES**

According to our expectations about Copernican Principle, Large Scale Symmetries and such, on *large scales* the geometry of the Universe should be maximally symmetric. The problem is that this is a *principle*, not a fact, and even if it is true, we cannot be sure, what is really *large scale*, what is *maximally* and what is *symmetric*.

In order to avoid pointless debates, let us accept that the Universe has a 4 (or, exceptionally, 5)-dimensional pseudo-Riemannian geometry

$$ds^2 = g_{ik}(x^l)dx^i dx^k \qquad (A.1)$$

where $x^1$, $x^2$ and $x^3$ are the spatial coordinates and $x^4=ct$ is the temporal one; henceforth with a convenient choice of units $c=1$ will be used.

Now the symmetry is generally understood as a Killing symmetry, so the existence of Killing vectors in whose directions the Lie derivatives of the geometry vanish, so infinitesimal triangles transported will remain geometrically unchanged (identical) (Eisenhart, 1950):

$$L_K(g_{ik}) = 0 \qquad (A.2)$$

Let us accept this definition for *symmetry*. Then in n=4 dimensions the maximal number of symmetries is n(n+1)/2=10. Such geometries are, however, contrary to observations (Robertson & Noonan 1969)

Without going into cosmological details, the first possibility for maximal *Killing* symmetries is maximal *spatial* symmetry, 6 Killing vectors, with SO(4), SO(3,1) or E(3) symmetries. Recent observations seem to favour E(3), which means even one "initial conditions" less than the others. We may accept this, and then (A.1) gets the specific form

$$ds^2 = dt^2 - R^2(t)\{dx^2+x^2(d\theta^2+\sin^2\theta d\varphi^2)\} \tag{A.3}$$

where R(t), the scale factor of the Universe, has no direct connection with the correlation quantities of the main text.

Then space is homogeneous and isotropic, so indeed there is no preferred point (and direction) in the Universe. However surely this spatially totally symmetric line element cannot be true for *small*, or local scales, such homogeneity being contrary to our own existence.

Inhomogeneities, hierarchies & such were not challenged up to galaxies in the early 20[th] century, and up to galaxy clusters in the middle 20[th]. Until cca. 1990, however, the existence of higher inhomogeneities/structures were doubtful.

This comes from the Big Bang Scenario. As earlier, as the Alpher-Bethe-Gamow picture (Wagoner, Fowler & Hoyle, 1967), proto-galaxy-clusters were consequences of decoupling radiation from massive matter at the primordial formation of neutral atoms at ~3000 K. Afterwards the decoupled photons cannot destroy density fluctuations and then gravity with positive feedback will enhance inhomogeneities (we refer Kämpfer, Lukács & Paál, 1995 as a monograph). That will be the end of *small scales* here, somewhere in the 10-50 Mpc range. Then, maybe, geometry becomes more and more homogeneous, going to *large scales*.

Then about 1990 lots of inhomogeneities were found on larger scales, called Large Walls and such (see e.g. Bellanger & de Lapparent 1995, Broadhurst & al. 1990). While the existences of these inhomogeneities are established facts, we do not have clear and obvious explanations *why* they appear; you may think about very early particle physical processes or such, as you like. These inhomogeneities surely appear at the high half of the 10 Mpc range and the low half of the 100 Mpc ones. Large Walls seem to appear at 128 or 140 Mpc separations (Broadhurst & al. 1990; Paál, Horváth & Lukács 1992 &c.) Gamma-bursts are anisotropic up to cca. 300 Mpc (Balázs, Mészáros & Horváth 1998).

Now, there is an alternative. Either the Universe is chaotic on *large scales*, or this inhomogeneity is restricted to *middle scales*, say, up to 300 Mpc (Mészáros, 1997), and at really *large scales* homogeneity appears.

We cannot tell what is correct. However the introduction of inhomogeneous middle scales uphold the existence of a Cosmologic Principle on large ones, so until no counterevidence is shown, we can remain with this solution.

We, however, note that there is another solution, even if it is not too convenient. Namely defining symmetries as existence *conformal* Killing vectors

$$L_K(g_{ik}) = \Phi(x^l)g_{ik} \tag{A.4}$$

the form (A.3) is compatible with the existence of the maximal number of conformal symmetries, which is n+1)*(n+2)/2=15 (Lukács & Mészáros, 1985). Also, e.g., Randall & Sudrum (1999) relatively recently suggested a 5-dimensional space-time, whose *spatial* 5[th] dimension is connected with particle physics, and space-time is highly inhomogeneous in $x^5$. While this space-time is not a cosmologic one with Killing symmetries, it is such with *conformal* ones (Lukács, 2000). Namely, in 5 dimensions the maximal *conformal* Killing symmetry means 21 ones, and the Randall-Sudrum line element has all of them.

## APPENDIX B: REDSHIFT AND COMOVING DISTANCE

Let us accept the geometry (A.3) for any reason. Now let us fill this spacetime with galaxies of velocities $u^i(x^l)$. Being the galaxies *small scale* objects, in principle they could move in any way. So we could write

$$u^i = u^i{}_{(ls)} + u^i{}_{(pec)} \tag{B.1}$$

where $u^i{}_{(ls)}$ must be homogeneous + isotropic, so

$$u^i = \delta^i{}_0 \tag{B.2}$$

while the peculiar velocity could be anything. However, according to observations, peculiar velocities remain in the order of 100 km/s (Peebles, 1980), so permitting hope in large scale symmetries.

Then ignore $u^i{}_{(pec)}$; with the *cosmologic* velocity field (B.2) we can get the redshifts z from other galaxies as

$$1+z_i = R_0/R(t_i) \tag{B.3}$$

where

$$x_i = \int_t^{t_0} R^{-1}(t')dt' \tag{B.4}$$

So the connection z(x) can be calculated if we know R(t) for any time, so in a definite cosmologic model.

## APPENDIX C: ON THE "APPARATUS-SENSITIVITY" AT LARGE SCALES

Our Milky Way Galaxy may be quite a good example of substantial spiral galaxies. The Local Group has only two of this size, we and M31 Andromeda. The smallest ones, as e.g. the Magellans, are lost even in not too deep surveys, and intermediate ones are continuously dropping out with growing z's.

Now for the substantial spirals we may take us and Andromeda as examples, rather large ones, so an absolute brightness M=-18. At a distance 100 Mpc the corresponding apparent brightness is then m=+17, still well observable.

However at 600 Mpc, double of the possible lowest scale of large scale homogeneity we are down at m=+20.9. While the largest objects themselves are still observable without problem, the determination of z is harder and harder for the smaller ones, so they are dropping out. Recently the technique seems rather unsure beyond m=21.5 (Waddington & al. 2000).

## APPENDIX D: ON HOMOGENEITY OF CRYSTALLS

"Locally" inhomogeneous matter in a "globally" homogeneous space-time has a simple analogy in solid-state physics. Consider a regular crystallic lattice of, e.g., a metal, say, a cubic lattice with distances of neighbours *l*. This distance is in the order of $10^{-8}$ cm. On this scale the matter is clearly inhomogeneour, as X-ray crystallograpy demonstrates it.

However on macroscopic scale this *l* is simply a material constant of the metal. In usual cases *l* is independent of position, so matter is homogeneous in macroscopic sense; but we can prepare circumstances when it is not. E.g. if we maintain a substantial temperature *gradient* within the metal then l will vary with that gradient *on macroscopic scale*.

However in this analogous example the two scales are quite disjoint. Unfortunately in cosmologic contexts quite substantial inhomogeneities roughly analogous to lattice structure ("bubbles", "voids", "walls", &c.) are well documented almost up to 150 Mpc (Broadhurst & al. 1990), (Horváth, Lukács & Paál 1992), homogeneity is hoped from 300 Mpc (Mészáros 1997), while few rich galaxy samples go effectively beyond 600-800 Mpc.

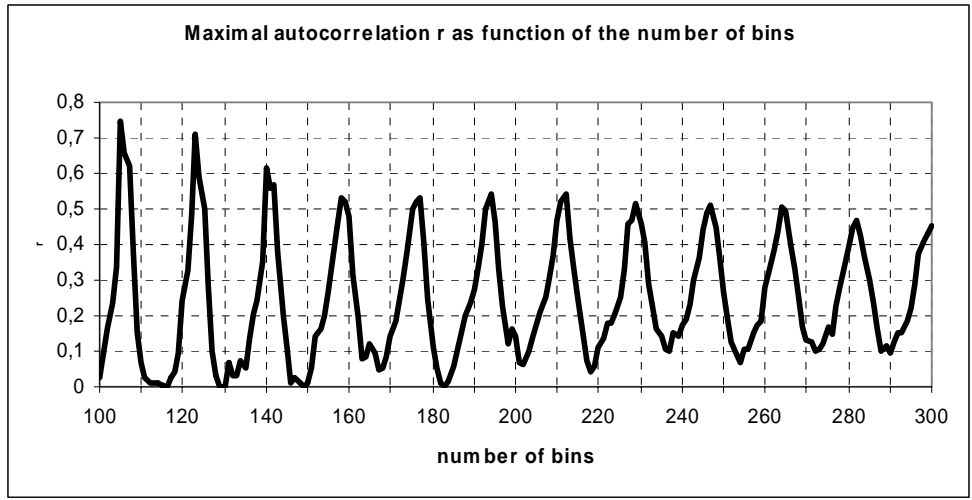

Fig. 5: The dependence of the autocorrelation coefficient (after steps in subchaps. 3.1-5) on bin number in a semi-random simulation sequence with relative frequency spread 0.10. In maxima the autocorrelation would seem highly significant, but in minima r is sometimes even lower than its fluctuation value (0.07 at 100 bins and 0.05 at 300.) According to Fig. 2, such frequency spreads generally result finally in cca. $R^2=5$. See Subchapt. 3.6, Chapt. 5 and App. E.

## APPENDIX E: REARRANGEMENT OF GALAXY NUMBER HISTOGRAMS INTO NEW BINS

Fig. 5 shows the consequences of such redistributions into new bins for a test sequence according to the "third level of randomness" mentioned in Chapter 5; with relative frequency spread 0.10. A clear quasiperiodic modulation of a moderately decreasing curve is seen. The quantity shown here is $R^2(v)$.

Now, the slow decrease comes surely from random fluctuations; as we saw even completely random sequences would give $R^2(v) \sim 1/v$. Here the signal is clear, since at peaks the curve is at 4-6 times the random level. Still, at some bin numbers the curve even goes *below* the level $1/v$.

We do not have to discuss the reason of this, since, as told in Subchap. 3.6, this behaviour is just the reason to average for bin numbers. However the curve gives us a lower bound, how much points are independent in the averaging. (Because of the Monte Carlo results of Chapter 5 we do not *need* this number; but order of magnitude agreement would be welcome.)

Obviously points of minima and maxima can be taken independent of each other. Now, Fig. 5 shows 12 clear minima and 12 clear maxima over the range $200 \leq v \leq 400$. So roughly each $8^{th}$ point is certainly independent now, in agreement what was said in Chapter 4.